\documentclass{ICRES}

\usepackage{subfigure}

\begin{document}

\title{Should XAI Nudge Human Decisions with Explanation Biasing?}

\author{Yosuke Fukuchi$^1$ and Seiji Yamada$^{2,3}$}

\address{%
$^1$ Tokyo Metropolitan University, Tokyo, 191-0065, Japan\\
$^2$ National Institute of Informatics, Tokyo, 101-8430, Japan\\
$^3$ The Graduate University for Advanced Studies (SOKENDAI), Kanagawa, 240-0193, Japan \\
E-mail: fukuchi@tmu.ac.jp}

\begin{abstract}
This paper reviews our previous trials of \textit{Nudge-XAI}, an approach that introduces automatic biases into explanations from explainable AIs (XAIs) with the aim of leading users to better decisions, and it discusses the benefits and challenges.
Nudge-XAI uses a user model that predicts the influence of providing an explanation or emphasizing it and attempts to guide users toward AI-suggested decisions without coercion.
The nudge design is expected to enhance the autonomy of users, reduce the risk associated with an AI making decisions without users' full agreement,
and enable users to avoid AI failures.
To discuss the potential of Nudge-XAI, this paper reports a post-hoc investigation of previous experimental results using cluster analysis.
The results demonstrate the diversity of user behavior in response to Nudge-XAI, which supports our aim of enhancing user autonomy.
However, it also highlights the challenge of users who distrust AI and falsely make decisions contrary to AI suggestions,
suggesting the need for personalized adjustment of the strength of nudges to make this approach work more generally.
\end{abstract}


\bodymatter

\section{Introduction}
The blackbox nature of machine learning models (particularly deep learning) makes it difficult for users to judge whether they should rely on an AI decision,
which can trigger ineffective use of AIs such as over-/under-reliance~\cite{doi:10.1518/001872097778543886}.
However, an increasing number of explainable AIs (XAIs) are actively being developed for this problem~\cite{8466590},
and they demonstrate the effectiveness of presenting explanations along with AI predictions in diverse applications~\cite{10.1145/3610218,electronics12214430,10.1145/3581641.3584055}.
The explanations include visual explanations such as saliency maps~\cite{zhang2018visual,6875957,samek2016evaluating} and linguistic explanations~\cite{VANDERWAA2021103404,10.1145/3448016.3458455}.
The development of large language models (LLMs) has also made it possible to generate various post-hoc explanations that support AI predictions\cite{wiegreffe-etal-2022-reframing}.

This paper discusses the design of XAIs that enables them to properly decide \textit{what to} or \textit{how to} explain rather than just thoughtlessly presenting all possible explanations.
Although having a variety of explanations could lead to multiple perspectives on AI predictions and deepen users' understanding,
imprudently providing all available explanations does not necessarily contribute to better decision-making or even result in undesirable outcomes.
For example, uninterpretability of explanations~\cite{maehigashi2023modeling,maehigashi2023roman,10.1145/3623809.3623834},
information overload~\cite{ferguson2022explanations,herm2023impact}, and contextual inaccuracy~\cite{10.1145/3301275.3302316} have harmful effects on users' cognitive load, 
task performance, and task time.

Studies related to this problem include comparisons of various explanation types or modalities to determine their effectiveness~\cite{NAISEH2023102941,herm2023impact}.
These findings provide engineers with guidelines for designing more appropriate and user-friendly explanations.
Another approach is developing an algorithm to automatically evaluate XAI explanations.
Wiegreffe et al. propose a method of evaluating explanations generated by LLMs by predicting human ratings of their acceptability~\cite{wiegreffe-etal-2022-reframing}.

This paper specifically emphasizes the importance of the dynamics and behavioral aspects of human-XAI interaction.
In interaction, a human evolves their understanding of a task and AI to figure out whether the AI is trustworthy and reliable compared with themselves~\cite{10.1371/journal.pone.0229132,9281021}.
This means that what information an XAI should provide constantly changes, and an XAI needs to be aware of these dynamics to properly communicate its reliability,
but most previous studies focus only on the static characteristic of explanations.
Automatic evaluation of explanations is promising regarding this point because it helps XAIs dynamically change what to explain.
However, the previous work considers users' perceptions of the explanations,
 and the behavioral aspect of human-XAI collaboration,
that is, how explanations provided affects the performance of decision-making
and how an XAI can proactively contribute to improving task performance by adaptively changing explanations, is neglected.

Driven by this motivation, this paper aims to deepen the discussion on our previously proposed Nudge-XAI~\cite{fukuchi2024selection,fukuchi2024emphasis},
which attempts to affect user decision-making by dynamically biasing how XAI explanations are presented.
Nudge-XAI algorithmically determines the manner in which explanations are presented---deciding whether to show an explanation~\cite{fukuchi2024selection} or whether to emphasize it~\cite{fukuchi2024emphasis}---using a user model that predicts user decisions for various scenarios regarding how explanations are provided.
Nudge-XAI aims to guide users to an AI-suggested decision through biasing and without coercing them into it.
The design is inspired by libertarian paternalism~\cite{b35e72fa-fff9-37d3-a508-45875042aa96},
the idea of leading humans to make better decisions while respecting their freedom of choice.
By integrating this concept, Nudge-XAI is also expected to reduce the risk associated with an AI making decisions without users' full agreement
and leave room for users to avoid AI failures.

This paper presents an analysis of prior experimental findings through cluster analysis to explore the potential of Nudge-XAI.
The findings reveal a wide range of user reactions to Nudge-XAI, aligning with our goal of improving user autonomy.
Nevertheless, the analysis also uncovers the issue of users who lack trust in AI, leading them to make decisions against the AI's suggestions.
This indicates the importance of proper personalization, paricularly, adjusting the strength of the nudges for each user in this approach.

\section{Nudge-XAI}
\subsection{Formalization}
In this paper, Nudge-XAI refers to an XAI system that aims to strategically guide users to a particular decision without coercing
but rather indirectly influencing them by biasing explanations. 
The design is inspired by libertarian paternalism~\cite{b35e72fa-fff9-37d3-a508-45875042aa96},
the idea of leading humans to make better decisions while respecting their freedom of choice.

Our previous studies proposed two implementations of Nudge-XAI.
X-Selector~\cite{fukuchi2024selection} is a method for selecting which explanations to provide,
and DynEmph determines whether to put emphasis on an explanation using a communication robot~\cite{fukuchi2024emphasis}.
They share the same concept of predicting the effect of such biasing on user decisions and guiding users to an AI-suggested decision by biasing.

Nudge-XAI is formalized with UserModel and $\pi$.
The former is a model of a user who makes a decision $d_u$ given explanations.
\begin{equation}
  \mathrm{UserModel}(\bm{c}, \bm{x}, d) = P(d_u = d| \bm{c}, \bm{x}).
\end{equation}
UserModel represents a probability distribution of $d_u$ conditioned by a set of explanations $\bm{x}$ and other contextual variables $\bm{c} = \{c_j\}$.
For DynEmph, let $\bm{x}$ be a set of tuple $(x_i, e_i)$, where $x_i$ is a sample of explanation, $i$ is the index for it, and $e_i$ is a Boolean that represents whether to emphasize $x_i$.
$\bm{c}$ includes other information such as AI predictions, task status, and user status. 
UserModel can be implemented with machine learning in a supervised manner using interaction logs.
In previous work, we obtained logs from crowdworkers' trials and used deep learning for implementing the model.

\begin{figure}[t]
  \begin{center}
      \includegraphics[width=0.7\linewidth]{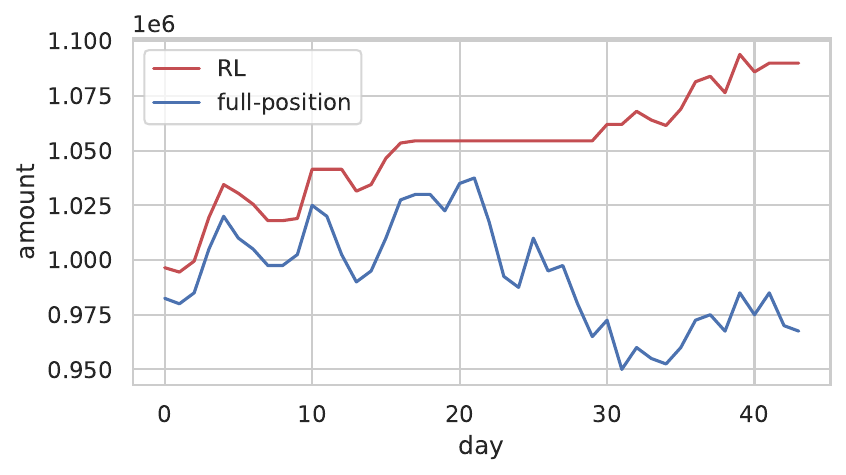}
    \caption{Performance of $\pi$ and result in which user was always took full position}\label{fig:rl_performance}
    \end{center}
\end{figure}

$\pi$ infers an AI-suggested decision $d_\mathrm{AI}$.
\begin{equation}
  \pi(\bm{c}, d) = P(d_\mathrm{AI} = d| \bm{c}).
\end{equation}
The inference is independent of user decision-making.
$\pi$ can be implemented with reinforcement learning.
Figure \ref{fig:rl_performance} illustrates the performance of $\pi$ used in our experiment compared with the full-position result in which user was always took full position.

Nudge-XAI explores which $\bm{x}$ will minimize the expected difference between $d_u$ and $d_\mathrm{AI}$ by simulating how $\bm{x}$ will affect $d_u$ with UserModel.
For DynEmph, the decision of whether to emphasize is calculated as:
\begin{equation}\label{eqn:distance} 
   \hat{\bm{x}} = \mathrm{argmin}_{\bm{x}} E_d[|\mathrm{UserModel}(\bm{c}, \bm{x}, d) - \pi(\bm{c}, d)|].
\end{equation}
If the user guidance of Nudge-XAI is perfect, a user's performance is expected to match the RL result (Fig. \ref{fig:rl_performance}).

\subsection{Implementation}
We evaluated X-Selector and DynEmph in a stock-trading simulator with the support of an XAI-based decision support system.
Hereafter, we explain the case of DynEmph.
In the simulation, participants were given virtual 1M JPY, checked the opening price and a price chart for each day, and decided whether to buy, sell, or hold their position with the decision support system.
Here, the support system showed its prediction of future stock prices and explanations for the prediction.
The prediction results were represented as the probability of bullish (over +2\%), neutral (-2\%--2\%), and bearish (under -2\%) predictions.
On the basis of the information, the participants input the day's order, and the simulator immediately transited to the next day.
Let $d$ be the amount of the position after the day's order.
This was repeated for 45 successive virtual days.

\begin{figure}[t]
  \begin{center}
      \includegraphics[width=0.7\linewidth]{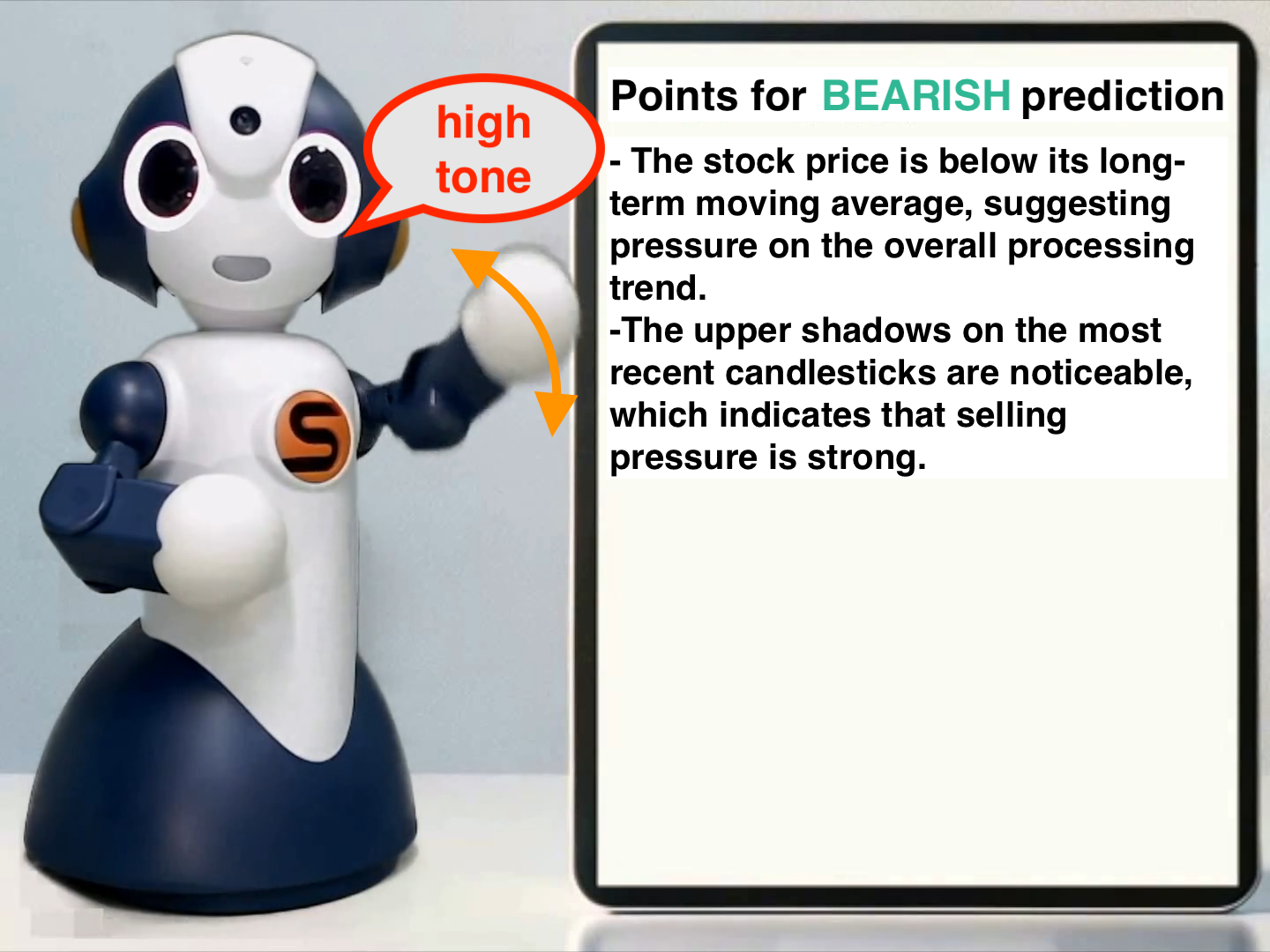}
    \caption{Emphasis expression with communication robot}\label{fig:sota}
  \end{center}
\end{figure}

Figure \ref{fig:sota} shows an example of an explanation.
The system has a set of natural language explanations generated with GPT-4V~\cite{openai2023gpt4} for the three prediction labels.
DynEmph decided whether to emphasize explanations for each label.
An expression of emphasis was implemented with a higher voice tone, repetitive swinging of the left arm, and lighting up of the LEDs on its eye contours.

$d_\mathrm{AI}$ was inferred by a reinforcement learning model.
In the simulation, $d_\mathrm{AI}$ was able to earn 90k JPY, which was better than most of the participants' results,
whereas a user consistently taking a full position from the beginning lost 32.5k JPY.

\section{Data analysis}
\subsection{Procedure}
To discuss the potential benefits and challenges of Nudge-XAI, we conducted a post-hoc analysis on experimental data for DynEmph~\cite{fukuchi2024emphasis},
where a communication robot automatically selected whether to emphasize explanations.

We conducted a cluster analysis of the data to discuss user behavior in detail.
Let $\vec{d_u}$ be a vector that represents the sequence of a user $u$'s 45 day's worth of decisions, that is, how much a user bought or sold stocks.
We first applied a PCA\footnote{Principal component analysis} to each $u$, and let us denote the transformed result as $\vec{d_u}'$.
We chose the first four principal components for $\vec{d_u}'$.
Then, we applied a k-means clustering for $\vec{d_u}'$, where $k = 4$, and acquired four clusters.

\subsection{Results}
\begin{figure}[t]
  \begin{center}
      \includegraphics[width=\linewidth]{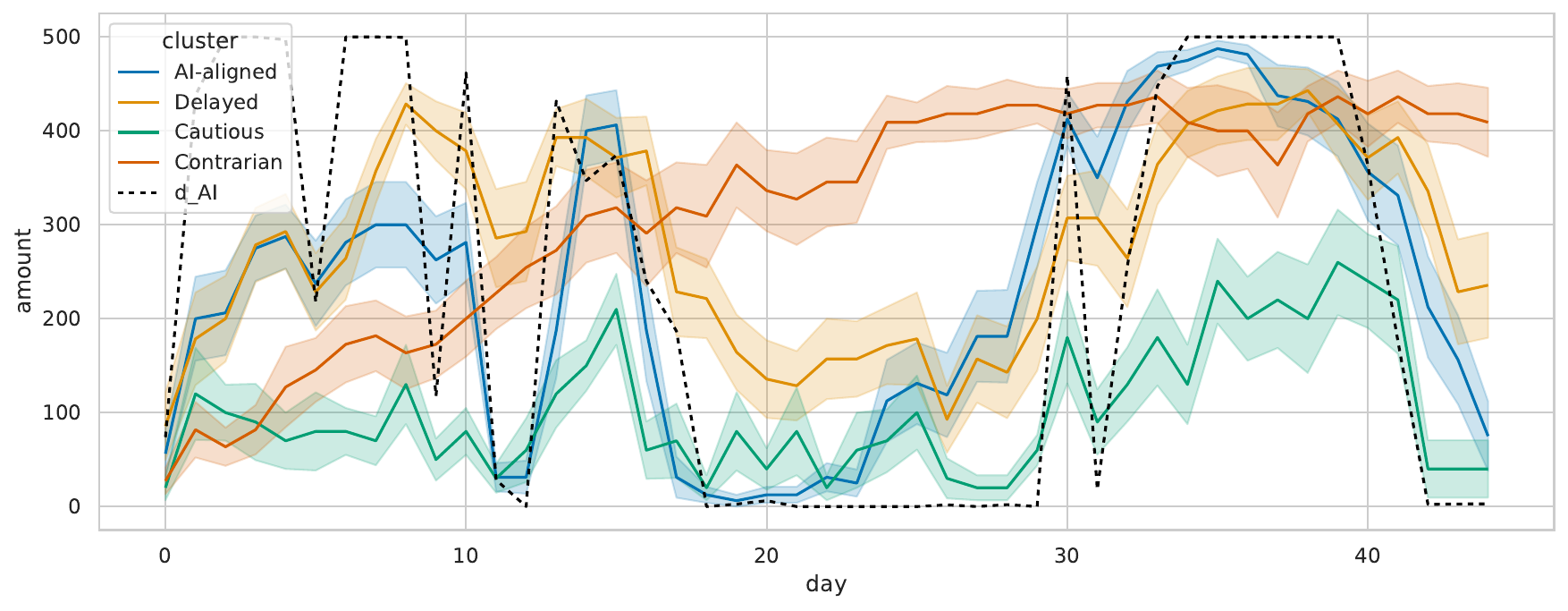}
    \caption{Averaged amount of positions for each cluster and $d_\mathrm{AI}$. Error bands show standard errors.}\label{fig:amount}
  \end{center}
\end{figure}

\begin{figure}[t]
  \begin{center}
      \includegraphics[width=0.5\linewidth]{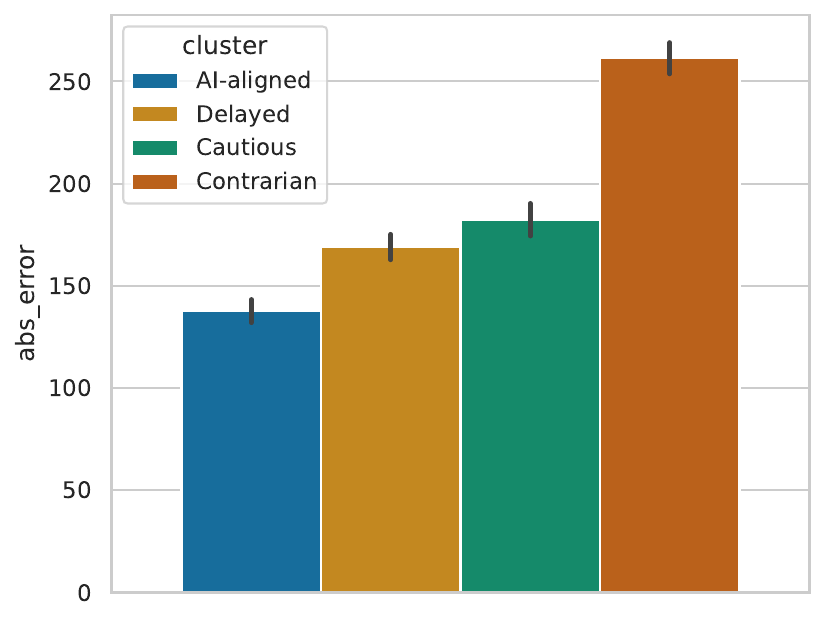}
    \caption{Absolute errors between $d_u$ and $d_\mathrm{AI}$}\label{fig:abs_error}
    \end{center}
\end{figure}

\begin{figure}[t]
  \begin{center}
    \subfigure[Distribution of correlation coefficients between $d_u$ and $d_\mathrm{AI}$]{
      \includegraphics[width=0.45\linewidth]{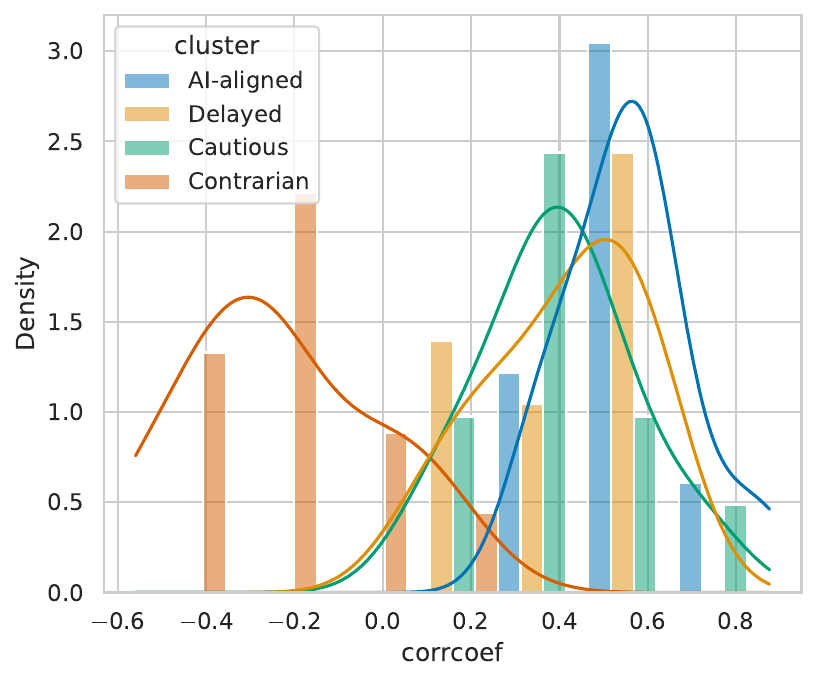}\label{fig:corrcoef}
    }
    \subfigure[Distribution of final assets]{
      \includegraphics[width=0.45\linewidth]{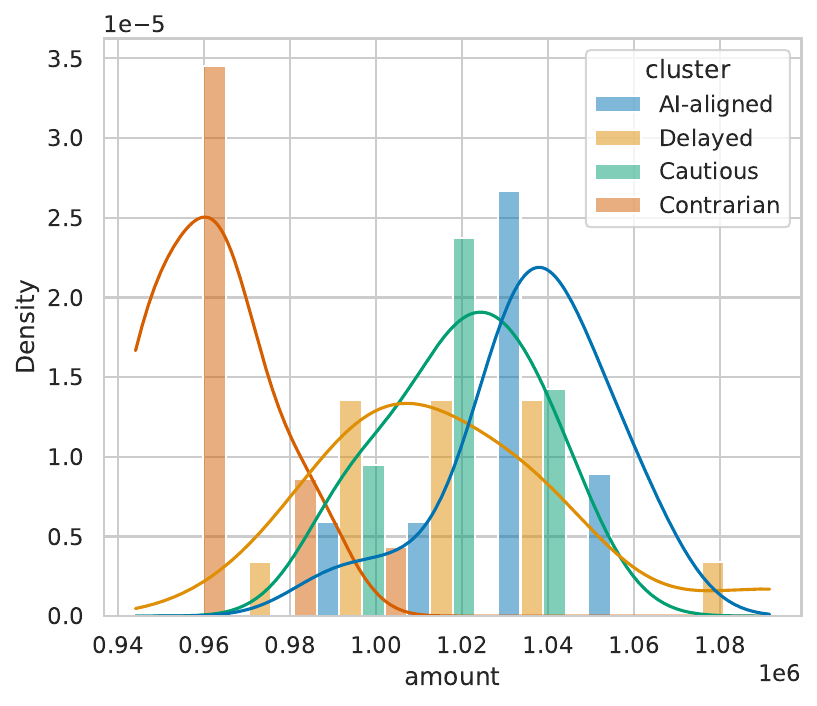}\label{fig:profit}
    }
    \caption{Distribution of correlation coefficients and final assets}\label{fig:high}
  \end{center}
\end{figure}

Figure \ref{fig:amount} shows $d_u$ averaged among participants for each cluster.
For comparison, the black dotted line shows $d_\mathrm{AI}$.
The result illustrates that DynEmph allowed for the diversity for $d_u$ while guiding the participants to $d_\mathrm{AI}$.
We investigated the characteristics of the four clusters and labeled them as AI-aligned, Delayed, Cautious, and Contrarian.
16, 14, 10, and 11 participants were assigned to them, respectively.
The clusters were in ascending order of the mean absolute differences between $d_u$ and $d_\mathrm{AI}$ (Figure \ref{fig:abs_error}).
The correlation coefficients between $d_u$ and $d_\mathrm{AI}$, which is a metric of DynEmph's successful guidance, were also in this order (Figure \ref{fig:corrcoef}),
but the order of their final performance, or the total amount of their assets, was reversed between Delayed and Cautious (Figure \ref{fig:profit}).
Let us take a closer look at each cluster in the order of their task performance.

\begin{figure}[t]
  \begin{center}
      \includegraphics[width=\linewidth]{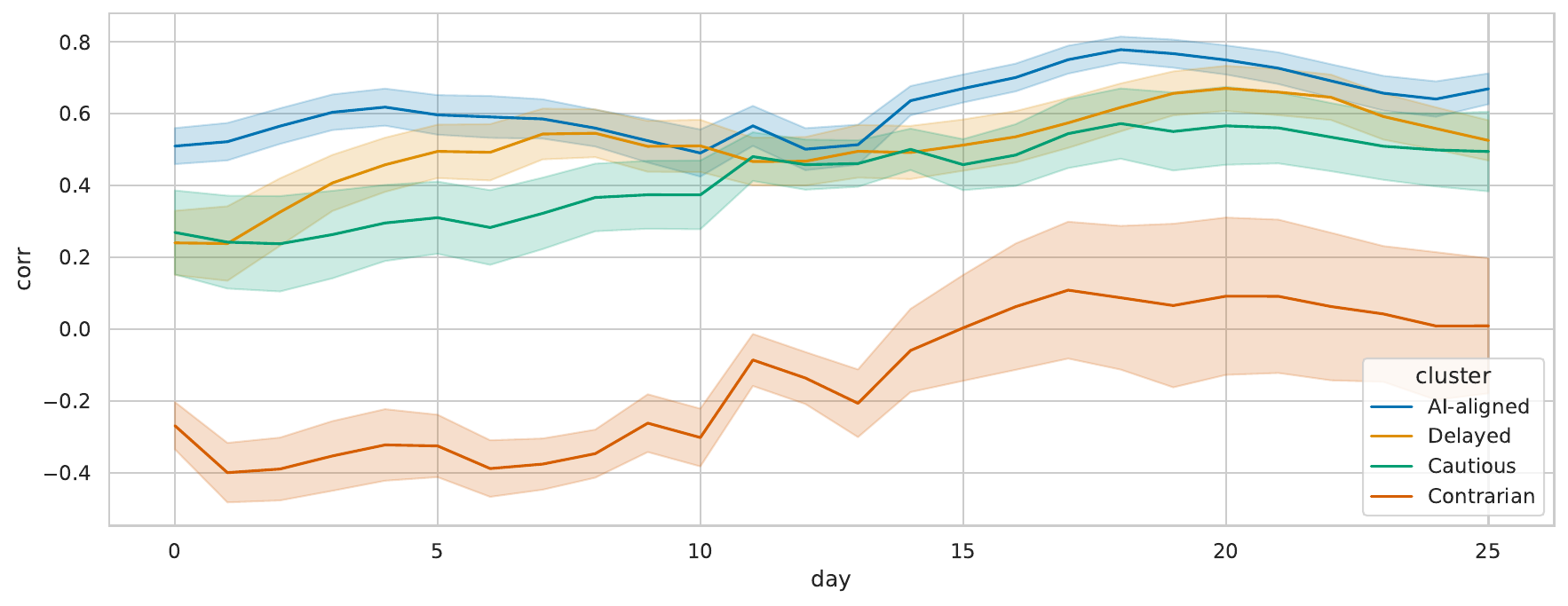}
    \caption{Moving average of correlation coefficients with window size 20}\label{fig:moving_corrcoef}
  \end{center}
\end{figure}

\subsubsection{AI-aligned}
The AI-aligned cluster is the group of the participants whom DynEmph guided to $d_\mathrm{AI}$ the best.
The correlation coefficients were constantly high throughout the trial (Figure \ref{fig:moving_corrcoef}).
As a result, the participants in this cluster earned more than those in the other clusters (Figure \ref{fig:profit}).
These were successful examples of DynEmph guiding users to better decisions without coercion.

\subsubsection{Cautious}
The Cautious participants were characterized by trading the least amount of stocks among the clusters (Figure \ref{fig:amount}).
They traded in spots and did not necessarily buy stocks when $d_\mathrm{AI}$ was high, which decreased the correlation coefficients,
but they achieved the second-highest profits among the clusters.
This is a positive case of Nudge-XAI embracing the autonomy of users.

\subsubsection{Delayed}
Although Delayed showed the second largest correlation coefficient, suggesting the influence of DynEmph's guidance,
this did not improve the performance (Figure \ref{fig:profit}).
The performance was inferior to Cautious despite the higher correlation.
Let us compare their trades with $d_\mathrm{AI}$ (Figure \ref{fig:amount}).
The Delayed participants were in moderate accordance with $d_\mathrm{AI}$
but failed to immediately follow the steep decrease of $d_\mathrm{AI}$ (days 10--12, 16--18), resulting in losing their assets.
This implies that the strength of the nudge was not enough for them,
which is a challenge for Nudge-XAI when a user needs to make an immediate decision.

\subsubsection{Contrarian}
Contrarian was the worst example in this trial.
Compared with the correlation coefficients of the other clusters, those of Contrarian was constantly low (Figure \ref{fig:moving_corrcoef}).
In the first half of the trial, the values were negative,
meaning that the participants in this cluster instead made decisions contrary to $d_\mathrm{AI}$.
While the existence of this cluster supports the aim of embracing users' freedom of choice with Nudge-XAI,
this did not result in better performance in this trial (Figure \ref{fig:profit}).
Therefore, when viewed in hindsight, the participants in this cluster are considered to have had an under-trust,
that is, they underestimated the performance of the decision support system and could not make use of it.

\section{Discussion}
\subsection{Did Nudge-XAI embrace freedom of choice?}
The diversity of $d_u$ among the clusters (Figure \ref{fig:amount}) demonstrates that collectively, Nudge-XAI did not coerce the participants into $d_\mathrm{AI}$ 
and successfully embraced the freedom of choice while biasing how explanation was provided. 
In addition, Figure \ref{fig:moving_corrcoef} shows moderate increases in the correlation coefficients, particularly for Contrarian,
indicating that the participants spontaneously increased their reliance on the decision support system in the interaction.
Therefore, we can conclude that Nudge-XAI embraced freedom of choice.
However, presumably, because the performance of $d_\mathrm{AI}$ was high enough in this trial, we did not observe a massive decrease of reliance,
so further experiments such as artificially decreasing the reliability of $d_\mathrm{AI}$ are required to investigate whether the design of Nudge-XAI can allow users to avoid AI failure by reserving freedom of choice.

\subsection{Did Nudge-XAI contribute to improving task performance?}
The effects differed among the clusters.
AI-aligned got the largest benefit from Nudge-XAI by obediently following the nudges, meaning that Nudge-XAI was effective for these users under the condition of the better performance of $d_\mathrm{AI}$.
The weak but positive correlation coefficients of Cautious suggest a certain influence of explanation biasing, and at least, we can say that Nudge-XAI was not harmful to the decisions of those users.
Nudge-XAI could not contribute to improving the task performance of Delayed and Contrarian as a result.
In particular, Contrarian's negative performance and negative correlation coefficients between $d_u$ and $d_\mathrm{AI}$ indicate a risk of Nudge-XAI being poorly trusted and falsely guiding some users to decisions contrary to AI suggestions.

\subsection{Should XAI nudge human decisions with explanation biasing?}
The results suggested several benefits of an XAI that nudges human decisions with explanation biasing.
Nudge-XAI actually contributed to improving the task performance of a part of users (AI-aligned) by guiding them to AI-suggested decisions.
Users can also make partial use of Nudge-XAI by exercising their freedom of choice (Cautious).
The diversity of $d_u$ implied the potential of enabling users to avoid AI failures as a form of collective intelligence.
The design is expected to reduce the risk associated with an AI making decisions without users' full agreement.

However, freedom of choice also caused a challenge for some users failing to make an immediate decision when necessary.
In addition, some users falsely made decisions contrary to AI suggestions, resulting in low performance.
These challenges need to be overcome to fully obtain the benefits of Nudge-XAI.
A solution can be to adjust the strength of the nudge depending on the characteristics of the users.
Nudge-XAI may be able to contribute to the Delayed participants by strengthening the nudge depending on the urgency of decision-making,
and on the other hand, weakening the nudge may work for Contrarian so as not to lead to false guidance.
This paper focused on the sole effect of explanation biasing, but Nudge-XAI could also integrate other communication modalities to repair trust for Contrarians and give a stronger message to Delayed.


\section{Conclusion}
This paper conducted a post-hoc analysis of experimental data for Nudge-XAI, 
an approach in which XAI explanations are automatically biased to lead users to better decisions,
to discuss its benefits and challenges.
The results suggested that our implementation of Nudge-XAI successfully enhanced the autonomy of user decision-making.
The guidance was beneficial for AI-aligned users who obediently followed the nudges,
but we also found some cases in which Nudge-XAI could not trigger users to make immediate decisions.
We also found cases in which Nudge-XAI falsely guided users to decisions contrary to AI-suggested ones because of their under-trust in the AI.
This indicates the importance of properly adjusting the strength of nudges depending on the characteristics of users.

\section*{Acknowledgments}
This work was supported in part by JST CREST Grant Number JPMJCR21D4 and JSPS KAKENHI Grant Number JP24K20846.

\bibliographystyle{IEEETran}
\bibliography{bib}

\end{document}